\documentclass[twoside,fleqn]{article}
\usepackage{espcrc2}
\usepackage{amssymb}
\usepackage[dvips]{epsfig}
\newcommand{\be}{\begin{equation}}
\newcommand{\ee}{\end{equation}}
\newcommand{\bea}{\begin{eqnarray}}
\newcommand{\eea}{\end{eqnarray}}
\newcommand{\bean}{\begin{eqnarray*}}
\newcommand{\eean}{\end{eqnarray*}}

\newcommand{\uts}[2]{#1^{{\scriptsize \mbox{#2}}}}

\newcommand{\cl}[1]{\mathcal{#1}}
\newcommand{\Tr}{\mbox{Tr\,}}
\newcommand{\tr}[1]{\mbox{Tr}\left(#1\right)}

\newcommand{\mv}[1]{\langle#1\rangle}
\newcommand{\mvl}[1]{\left\langle#1\right\rangle}

\newcommand{\half}{{\textstyle{\frac{1}{2}}}}

\newcommand{\ltz}{L_{{\scriptsize \mbox{z}}}}
\newcommand{\ltxy}{L_{{\scriptsize \mbox{x,y}}}}

\newcommand{\z}{\mbox{z}}

\newcommand{\lphi}{L_{\varphi}}

\newcommand{\plaqix}{U_{{\scriptsize \mbox{p}_i},x}}
\newcommand{\plaqsx}{U_{{\scriptsize \mbox{p}_s},x}}
\newcommand{\plaqtx}{U_{{\scriptsize \mbox{p}_t},x}}
\newcommand{\mh}{m_H}
\newcommand{\mw}{m_W}

\newcommand{\hw}{R_{HW}}

\newcommand{\bs}{\beta_s}
\newcommand{\bt}{\beta_t}
\newcommand{\ks}{\kappa_s}
\newcommand{\kt}{\kappa_t} 
\newcommand{\gb}{\gamma_{\beta}}
\newcommand{\gk}{\gamma_{\kappa}}

\newcommand{\hf}[1]{#1_{x}}
\newcommand{\hfup}[2]{#1_{x+\hat{#2}}}

\newcommand{\gf}[2]{#1_{x,#2}}

\newcommand{\lkop}[3]{#1_{#2;#3}}
\title{
{
\vspace{-4.5cm} \normalsize \hfill
\parbox{30mm}{MS-TPI-97-11\\ITP-Budapest 535\\KEK-TH-541\\
hep-lat/9709098}
}\\[30mm]
The electroweak phase transition at $\mh\simeq 80$ GeV from
$L_t=2$ lattices%
\thanks{Poster presented by J.\ H.\ at LATTICE~97, Edinburgh.}
}
\author{
Ferenc Csikor\address{
Institute for Theoretical Physics, E\"otv\"os University,
H-1088 Budapest, Hungary},
Zoltan Fodor\address{
KEK Theory Group, Tsukuba-shi, Ibaraki 305, Japan}%
\thanks{On leave from Institute for Theoretical Physics,
E\"otv\"os University, H-1088 Budapest, Hungary.}
and Jochen Heitger\address{
Institut f\"ur Theoretische Physik I, Universit\"at M\"unster,
D-48149 M\"unster, Germany}
}
\begin{document}
%
%%%%%    slight change in table style, J.H. 1996)    %%%%
\makeatletter
\long\def\@maketablecaption#1#2{\vskip 10mm #1. #2\par}
\makeatother
%%%%%%%%%%%%%%%%%%%%%%%%%%%%%%%%%%%%%%%%%%%%%%%%%%%%%%%%%
%
\begin{abstract}
We study the finite-temperature electroweak phase transition by numerical
simulations of the four-dimensional SU(2)-Higgs model on anisotropic
lattices with temporal extension $L_t=2$.
The physically interesting parameter region of Higgs masses near 80 GeV
is reached, and recent results on some thermodynamic quantities are
presented.
\end{abstract}
\maketitle
%
%%%%%%%%%%%%%%%%%%%%%%%%%%%%%%%%%%%%%%%%%%%%%%%%%%%%%%%%%%%%%%%%%%%%%%%%%
%
\section{Introduction}
Though the scenario of electroweak baryogenesis at a sufficiently
strong first order phase transition \cite{K72KL7276KRS85S87} within the
SM seems to be ruled out \cite{KLRS9697KNPR97GIS97,FHJJM95CFHJM96}, it
appears important to quantify its nature and strength at more realistic
Higgs masses $\mh\simeq 80$ GeV, and to compare with effective
$3D$--theories claiming an endpoint of the transition line at
$\uts{\mh}{(crit)}\lesssim 80$ GeV, beyond which the EWPT turns into an
analytic crossover \cite{KLRS9697KNPR97GIS97}.
Hence we made numerical simulations of the anisotropic SU(2)--Higgs
model, since for weaker transitions at larger $\mh$ one expects the
typical excitations $m\ll T$ to require isotropic lattices exceeding most
accessible computer resources.

In the following we will focus on interface tension and latent heat from
$T>0$ simulations at $L_t=2\ll\ltxy\ll\ltz$.
These, involving a sequence of heatbath and overrelaxation algorithms,
were done at HLRZ, J\"ulich (CRAY-T90), and DESY-IfH, Zeuthen
(APE-Quadrics), Germany.
\section{Anisotropic SU(2)--Higgs model}
The lattice action of the four-dimensional SU(2)--Higgs model on
anisotropic lattices reads
\bea
&   &\hspace{-0.75cm}S[U,\varphi]=\sum_{x}\bigg\{\sum_{i=s,t}
     \beta_i\sum_{{\scriptsize \mbox{p}_i}}
     \Big(1-\half\,\Tr\plaqix\Big)\nonumber\\ 
&   &\hspace{-0.75cm}-\ks\sum_{\mu=1}^{3}\tr{\hfup{\varphi}{\mu}^+
     \gf{U}{\mu}\,\hf{\varphi}}-\kt\tr{\hfup{\varphi}{4}^+\gf{U}{4}\,
     \hf{\varphi}}\nonumber\\
&   &\hspace{-0.75cm}+\half\,\tr{\hf{\varphi}^+
     \hf{\varphi}}+\lambda\left[\,\half\,\tr{\hf{\varphi}^+
     \hf{\varphi}}-1\,\right]^2\bigg\}
\label{latact}
\eea
%
%%% Beginn Figur %%%
\begin{figure}[htb]
\begin{center}
\vspace{-0.5cm}
\epsfig{file=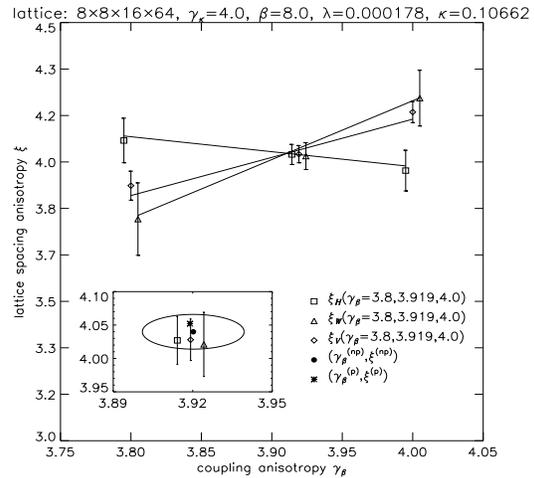,
             width=7.0cm,height=6.5cm}
\parbox{7.5cm}{
\vspace{-0.9cm}
\caption{\label{XiPlot} \sl $\xi$--evaluation at $\mh=72(5)$ GeV and
                            $g_R^2=0.577(15)$, whose equal abscissas are
                            displaced.
                            $\xi_i(\gb)$, as $s$-- to $t$--like
                            $m_i$--ratios in the Higgs and vector
                            channels ($i=H,W$) and from a suitable
                            mapping of static potentials ($i=V$), are
                            interpolated to coincide in errors.
                            The error ellipse of the matching point 
                            encloses the numerical estimates and the
                            perturbative one \cite{CF96CFH97}.}
}
\vspace{-1.2cm}
\end{center}
\end{figure}
%%% Ende Figur %%%
%
in terms of gauge links $\gf{U}{\mu}\in\mbox{SU(2)}$, space- and
timelike plaquettes $\plaqsx$ and $\plaqtx$, site variables
$\hf{\varphi}=\hf{\rho}\hf{\alpha}$, $\hf{\rho}>0$, 
$\hf{\alpha}\in\mbox{SU(2)}$, and the lattice spacing and coupling
anisotropy parameters $\xi\equiv a_s/a_t$, $\gb\equiv\sqrt{\bt/\bs}$,
and $\gk\equiv\sqrt{\kt/\ks}$ with $\beta^2=\bs\bt$ and $\kappa^2=\ks\kt$.
The general strategy of $T>0$ studies in $D=4$ is to fix
$T_c=1/a_tL_t$ at a given temporal extension $L_t$, to determine the
critical hopping parameter $\kappa_c$, and to calculate in this phase
transition point the physical, non-perturbatively renormalized parameters
$\hw\equiv\mh/\mw$ ($\uts{\mw}{(phys)}=80$ GeV) and $g_R^2$ in simulations
on $T=0$ lattices \cite{FHJJM95CFHJM96}.
Then the continuum limit is realized as approach to the scaling region via
$L_t\rightarrow\infty$ along lines of constant physics.

In \cite{CF96CFH97} we confirmed the one-loop corrections
$\uts{\gb}{(p)}=3.919$ and $\uts{\xi}{(p)}=4.052$ to the tree-level
anisotropies $\gb=\gk\equiv\xi\equiv 4$ non-perturbatively by demanding
space-time symmetry restoration (rotational invariance) with
correlation lengths in physical units being equal in both directions,
see figure~\ref{XiPlot}.
This opens the way to analyze the EWPT for $\mh\gtrsim 80$ GeV within the
$4D$--model in a systematic and fully controllable way. 
\section{Thermodynamic quantities and results}
\begin{figure}[htb]
\begin{center}
\vspace{-0.5cm}
\epsfig{file=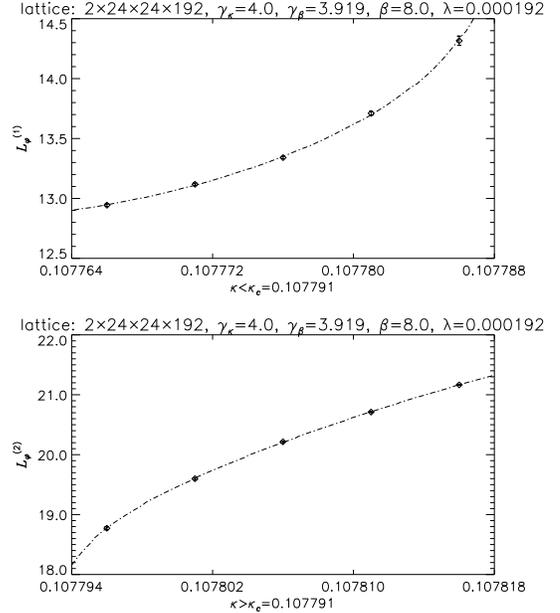,width=7.0cm}
\parbox{7.5cm}{
\vspace{-0.9cm}
\caption{\label{LphiPlot} \sl Four-parameter $\chi^2$--fit of
                              $\lphi^{(i)}$, $i=1,2$.}
}
\vspace{-1.2cm}
\end{center}
\end{figure}

\begin{figure}[htb]
\begin{center}
\epsfig{file=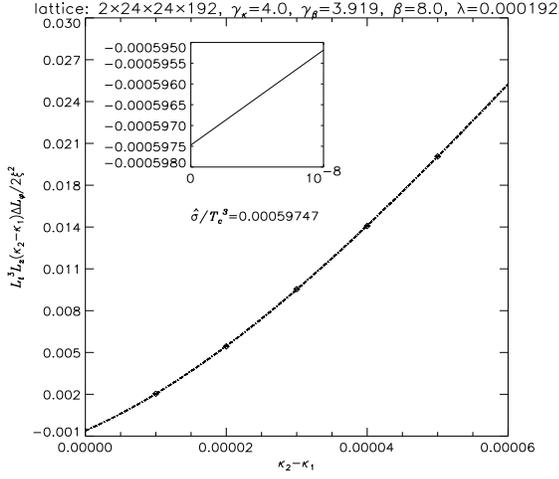,width=7.0cm,height=6.5cm}
\parbox{7.5cm}{
\vspace{-0.9cm}
\caption{\label{DLphiPlot} \sl As in figure~\ref{LphiPlot} but for
                               $\Delta\lphi\equiv
                               \lphi^{(2)}-\lphi^{(1)}$.}
}
\vspace{-1.2cm}
\end{center}
\end{figure}

In view of the large lattices to be used, the interface tension $\sigma$
has been determined by employing the two-coupling method 
\cite{PR89HPRS9091} in $\kappa$, which in previous investigations of the
SU(2)--Higgs model \cite{CFHH95HH96} turned out to be quite robust and,
at the same time, most economic among the other methods at disposal
\cite{B8182JMMTW89GIS97}.
After enforcing an interface pair perpendicular to the $\z$--direction
by dividing the lattice volume in symmetric and Higgs phases with
$(\kappa_1<\kappa_c:\z\le\ltz/2\,,\,\kappa_2>\kappa_c:\z>\ltz/2)$,
the related additional free energy $\Delta F$ yields for
$\Delta\kappa\equiv\kappa_2-\kappa_1\ll1$ the estimator
\cite{FHJJM95CFHJM96,CFHH95HH96}
\be
a_s^2a_t\sigma=
\frac{1}{2}\,\lim_{\Delta\kappa\rightarrow 0}\,\Big\{
\Delta\kappa\cdot\ltz\cdot\big[\,L^{(1)}_{\varphi}-L^{(2)}_{\varphi}
\,\big]\Big\}\,.
\label{sigma}
\ee
$L_{\varphi}^{(i)}=L_{\varphi}^{(i)}(\kappa_1,\kappa_2)$ denotes the
expectation value of the $\varphi$--link operator
$\lkop{L}{\varphi}{x\mu}\equiv
\half\Tr(\hfup{\varphi}{\mu}^+\gf{U}{\mu}\hf{\varphi})$ in the respective
phases, and, since $\Delta F\simeq\cl{O}(\Delta\kappa)$, the 
$(N+2)$--parametric Laurent ans\"atze
\be
L_{\varphi}^{(i)}=
-\frac{c_i}{\kappa_i-\kappa_c}
+\sum_{j=0}^{N}\gamma^{(j)}_i(\kappa_i-\kappa_c)^j+\,\cdots
\label{laurent}
\ee
give $\hat{\sigma}/T_c^3=L_t^3\ltz(c_1+c_2)/\xi^2$.
As exemplarily displayed in figures~\ref{LphiPlot} and \ref{DLphiPlot}, we
performed such fits to sets of 2--$\kappa$ data at $L_t=2$ with simulation
parameters corresponding to $\mh=78(4)$ GeV pole mass and
$g_R^2=0.539(16)$.
%
%%% Beginn Figur %%%
%%% Ende Figur %%%
%
%
%%% Beginn Figur %%%
%%% Ende Figur %%%
%
For the best fit we found $\hat{\sigma}/T_c^3=0.0006(3)$ on a lattice of 
size $2\times24^2\times192$ with $\chi^2/\mbox{dof}\simeq 1$, in
complete agreement with some data from a larger spatial volume.
When inspecting various fits along the available $\kappa$--intervals with
a reasonable number of fit parameters $\gamma^{(j)}_i$ in
eq.~(\ref{laurent}), the combined number quoted in table~\ref{ResTab}
covers the total spread of all reliable fit results, whose individual
errors include the statistical error from a bootstrap analysis and the
significant uncertainty in $\kappa_c$ \cite{CFHH95HH96}.

\begin{figure}[htb]
\begin{center}
\epsfig{file=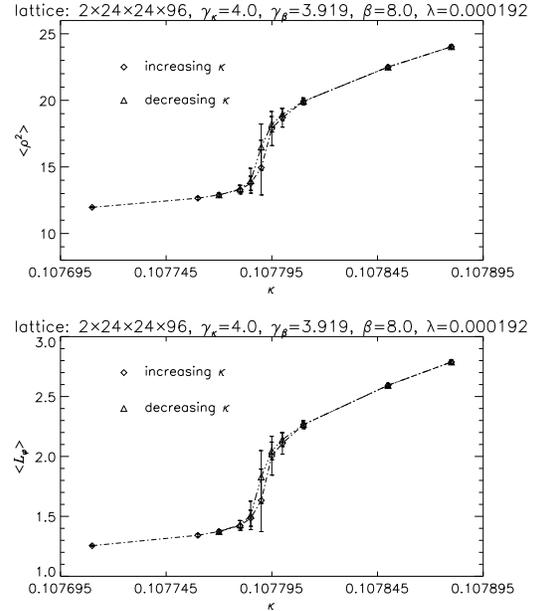,width=7.0cm}
\parbox{7.5cm}{
\vspace{-0.9cm}
\caption{\label{HystPlot} \sl Hystereses of the operators $\rho^2$ and 
                              $\lphi$ around the critical
                              $\kappa$--region at $L_t=2$.}
}
\vspace{-1.2cm}
\end{center}
\end{figure}

From quadratic fits of the discontinuities of the order parameters
showing up in the thermal cycles of figure~\ref{HystPlot}
%
%%% Beginn Figur %%%
%%% Ende Figur %%%
%
we also extracted the jump in the Higgs field vacuum expectation value,
here as $\Delta v/T_c=L_t\xi^{-1}\sqrt{2\kappa\,\Delta\mv{\rho^2}}$, and
in $L_t^4\xi^{-3}\,\frac{\partial\kappa}{\partial\tau}\,
\Delta\mvl{\lphi}$, $\tau\equiv-\ln(a_t\mw)$, which is the dominating
contribution to the latent heat if defined as the energy density
difference $\Delta\epsilon/T_c^4$ \cite{FHJJM95CFHJM96}.
Their numerical outcomes at identical parameters are collected in
table~\ref{ResTab} as well.
%
%%% Beginn Tabelle (einspaltig) %%%
\begin{table}[htb]
\begin{center}
\begin{tabular*}{7.5cm}{ccccc}
\hline
%   $L_t$ & $T_c/\mh$ & $\hat{\sigma}/T_c^3$ & $\Delta v/T_c$ 
  $L_t$ & $T_c/\mh$ & $10^4\hat{\sigma}/T_c^3$ & $\Delta v/T_c$ 
& $10^4\Delta\epsilon/T_c^4$ \\
\hline\hline
%   2 & 1.86(2) & 0.0006(4) & 0.37(16) & 0.0033(27) \\
  2 & 1.86(2) & 6(4) & 0.37(16) & 33(27) \\
  3 & 1.8(2)  & ---  & ---      & ---    \\ 
\hline
\end{tabular*}
\parbox{7.5cm}{
\vspace{-0.7cm}
\caption{\label{ResTab} \sl Lattice results at $L_t=2$ and,
                            preliminarily, at $L_t=3$.
                            The transition points lie at
                            $\kappa_c=0.107791(3)$ and
                            $\kappa_c=0.10703(3)$.}
}
\vspace{-1.2cm}
\end{center}
\end{table}
%%% Ende Tabelle %%%
%
\section{Discussion and outlook}
$\hat{\sigma}/T_c^3$ and $\Delta\epsilon/T_c^4$ for $\mh\simeq 80$ GeV
are substantially smaller than perturbatively ($\sigma/T_c^3\simeq 0.002$
\cite{FH94BFH9495}).
They are even consistent with a no first order phase transition scenario
approximately on the 1--$\sigma$ level.
The fact that this result deviates from those of the $3D$--investigations
\cite{KLRS9697KNPR97GIS97} should be clarified in future.
However, a temporal lattice extension of $L_t=2$ may be still too far
from continuum physics, and at least the knowledge of the behaviour at
$L_t=3$ seems necessary to draw a final conclusion.
%
%%%%%%%%%%%%%%%%%%%%%%%%%%%%%%%%%%%%%%%%%%%%%%%%%%%%%%%%%%%%%%%%%%%%%%%%%
%
% bibliography
%

%
\end{document}